\documentclass[electronic]{vgtc}             % electronic version

\ifpdf%                                % if we use pdflatex
  \pdfoutput=1\relax                   % create PDFs from pdfLaTeX
  \usepackage{graphicx}                % allow us to embed graphics files
  \DeclareGraphicsExtensions{.pdf,.png,.jpg,.jpeg} % for pdflatex we expect .pdf, .png, or .jpg files
\else%                                 % else we use pure latex
  \usepackage{graphicx}                % allow us to embed graphics files
  \DeclareGraphicsExtensions{.eps}     % for pure latex we expect eps files
\fi%

\graphicspath{{figures/}{pictures/}{images/}{./}} % where to search for the images

\usepackage{microtype}                 % use micro-typography (slightly more compact, better to read)
\PassOptionsToPackage{warn}{textcomp}  % to address font issues with \textrightarrow
\usepackage{textcomp}                  % use better special symbols
\usepackage{mathptmx}                  % use matching math font
\usepackage{times}                     % we use Times as the main font
\usepackage{cite}                      % needed to automatically sort the references
\usepackage{tabu}                      % only used for the table example
\usepackage{booktabs}                  % only used for the table example
\onlineid{1088}

\vgtccategory{System or Tool}

\vgtcinsertpkg

\usepackage{amsmath}

\DeclareMathOperator{\tf}{tf}

\DeclareMathOperator{\idf}{idf}

\DeclareMathOperator{\tfidf}{tf-idf}

\newcommand{\citeneeded}[1]{[?]}
\newcommand{\refneeded}[1]{[?]}

\usepackage{svg}
\usepackage{calc}
\svgsetup{inkscapelatex=false}

\newcommand{\toolname}[0]{\textsc{Almanac}}

\author{Terrell Ibanez\thanks{e-mail: terrell.ibanez@stanford.edu}\\ %
        \parbox{1.4in}{\scriptsize \centering Stanford University \\ Tableau Research} %
\and Vidya Setlur\thanks{e-mail: vsetlur@tableau.com}\\ %
     \scriptsize Tableau Research %
\and Maneesh Agrawala \thanks{e-mail: maneesh@cs.stanford.edu}\\ %
     \scriptsize \centering Stanford University}

\abstract{
Authors often add text annotations to charts to provide additional context for visually prominent features such as peaks, valleys, and trends. 
However, writing annotations that provide contextual information, such as descriptions of temporal events, often requires considerable manual effort. To address this problem, we introduce \toolname, a JavaScript API that recommends annotations sourced from the New York Times Archive of news headlines. \toolname{} consists of two independent parts: (1) a prominence feature detector and (2) a contextual annotation recommender. We demonstrate the utility of the API using D3.js and Vega-Lite to annotate a variety of time-series charts covering many different data domains. Preliminary user feedback shows that \toolname~is useful to support the authoring of charts with more descriptive annotations.

}

\begin{document}
\teaser{
\vspace{-1em}
    \includegraphics[width=\textwidth]{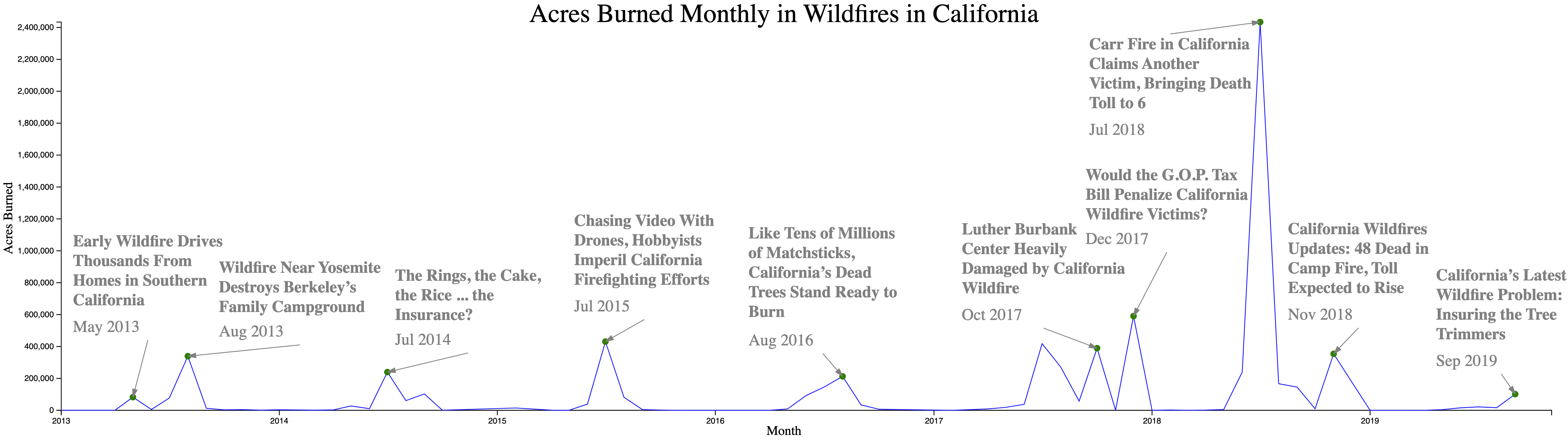}
    %\includesvg[inkscapelatex=false, width=\columnwidth]{figures/resultsV2/svg/wildfires.svg}
    \vspace{-2em}
    \caption{A time-series chart depicting the number of acres burned in various California Wildfires between 2013 and 2020. We used the \toolname~API to identify the prominent visual features of this chart (i.e. the peaks) and then annotate them with headlines from the New York Times. 
The headlines correctly name the fire responsible for each labeled peak and link to the corresponding article, thereby providing explanatory context for the peak. 
    In this case, we used the top-ranked headline identified by \toolname{}, but chart authors can manually choose lower-ranked headlines as well. The one peak left unlabeled did not return any suitable headline from the New York Times Archive because they did not cover the wildfire responsible for it.}% \maneesh{Should probably make a 1 column figure version to save space in the short paper.}}
    \vspace{1em}
    \label{fig:teaser}
}

%%
%% The "title" command has an optional parameter,
%% allowing the author to define a "short title" to be used in page headers.
\title{\toolname{}: An API for Recommending Text Annotations For Time-Series Charts Using News Headlines }
\maketitle

%%
%% The code below is generated by the tool at http://dl.acm.org/ccs.cfm.
%% Please copy and paste the code instead of the example below.
%%
% \begin{CCSXML}
% <ccs2012>
%    <concept>
%        <concept_id>10003120.10003145.10003151</concept_id>
%        <concept_desc>Human-centered computing~Visualization systems and tools</concept_desc>
%        <concept_significance>500</concept_significance>
%        </concept>
%  </ccs2012>
% \end{CCSXML}

% \ccsdesc[500]{Human-centered computing~Visualization systems and tools}

% \printccsdesc

\CCScatlist{
  \CCScatTwelve{Human-centered computing}{Visu\-al\-iza\-tion}{Visu\-al\-iza\-tion techniques}{Treemaps};
  \CCScatTwelve{Human-centered computing}{Visu\-al\-iza\-tion}{Visualization design and evaluation methods}{}
}

%%
%% Keywords. The author(s) should pick words that accurately describe
%% the work being presented. Separate the keywords with commas.

%%
%% This command processes the author and affiliation and title
%% information and builds the first part of the formatted document.

\section{Introduction}
% \begin{figure}[ht]
% \vspace{-1em}
%     %\includegraphics[width=\columnwidth]{figures/resultsV2/png/wildfires.png}
%     \includegraphics[width=\columnwidth]{figures/resultsV3/png/wildfires.png}
%     %\def\svgscale{1.5}
%     %\includesvg[width=\linewidth]{figures/resultsV2/svg/wildfires.svg}
%     \vspace{-2em}
%     \caption{A time-series chart depicting the number of acres burned in various California Wildfires between 2013 and 2020. We used the \toolname~ API to identify the prominent visual features of this chart (i.e. the peaks) and then annotate them with headlines from the New York Times. 
% The headlines correctly name the fire responsible for each labeled peak and link to the corresponding article, thereby providing explanatory context for the peak. 
%     In this case we used the top-ranked headline identified by \toolname{}, but chart authors can manually choose lower-ranked headlines as well. The one peak left unlabeled did not return any suitable headline from the New York Times Archive because they did not cover the wildfire responsible for it.} %\maneesh{Should probably make a 1 column figure version to save space in the short paper.}}
%     \vspace{1em}
%     \label{fig:teaser}
% \end{figure}

% Motivate the problem /
Text annotations play an important role in communicating insights about the data depicted in charts. 
Chart authors often include annotations to draw the readers’ attention to specific visual features (e.g., peaks, valleys, and trends) and to provide additional context~\cite{Lundgard2022AccessibleVV,Ren2017ChartAccentAF}. %Textual descriptions along with the visual features in the charts, have shown to be %influential in readers' takeaways. 
Such text is especially effective when it provides external information not directly present in the chart that explains the visually prominent features~\cite{Kim2021}. For example, Figure~\ref{fig:teaser} shows a time series of acres burned in California wildfires from 2013 to 2020~\cite{Ares2020Wildfires}, and the text annotations name the fire responsible for each peak while providing additional information about the damage they caused.

Yet, authoring such contextual annotations for the visually prominent features in a chart requires considerable manual effort. Authors have to identify the prominent features and examine a variety of external sources (e.g., news websites) to find relevant information. While researchers have developed prototype tools for automating chart annotation, the implementations have been fairly limited. For example, Contextifier~\cite{Hullman2013} is limited to a set of 11 stocks and corresponding financial news articles for the years 2010-2012, limiting its applicability to more general data sets.
%Moreover, as a research prototype, it is not designed to fit into the workflow of web-based chart designers. 

In this work, we present \toolname{}, a JavaScript API for recommending contextual annotations from the New York Times (NYT) Archive~\cite{NYTAPI}, to any time-series data visualization that is designed to fit into the workflow of web-based chart design. \toolname{}~provides two main components; {\bfseries \em (1) A prominence feature detector} that identifies visually prominent peaks, valleys, and trends in an input time-series chart.
While our built-in detector uses a topological persistence algorithm~\cite{Huber2020} to identify these features, the API is designed so that users can easily swap in other detectors or manually specify the prominent features to be annotated. {\bfseries \em (2) A contextual annotation recommender} that takes as input a time-series chart and a visually prominent feature (specified as a point or time range) and suggests contextually relevant headlines. We demonstrate the utility of \toolname{}~by annotating a variety of time-series data sets in D3.js~\cite{Bostock2011}, Vega-Lite~\cite{Satyanarayan2017}, Observable Notebooks~\cite{Observable} and Tableau~\cite{tableauextension}.

\section{Related Work}
\begin{figure*}[ht]
 %\centering
    \includegraphics[width=\linewidth]{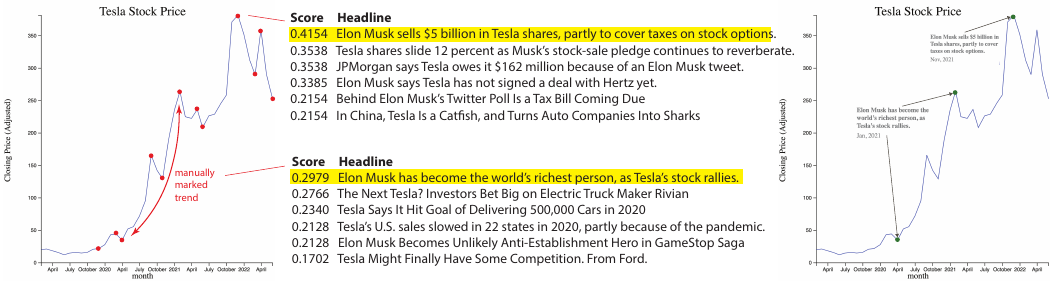}
    \vspace{-2em}
    \caption{Chart authors use the \toolname{} API in two steps. Given a time-series chart (left) of stock price data for Tesla Inc.~\cite{AlphaVantage2022Stock}, the author applies the {\ttfamily getChartFeatures} command to return the most prominent peaks and valleys in the chart (left, marked in red). Authors can manually mark point and trend features as well. The author then selects the most prominent peak and runs \toolname's {\ttfamily getAnnotations} command to obtain a  set of contextual headlines (middle top). By default, the highest-ranked headline is chosen as the annotation, but the author can manually examine the other headlines to see if a better headline is available. The author uses the top-ranked default headlines to create the final chart (right).}
    \vspace{-1em}   
    \label{fig:usingAPI}
\end{figure*}

%Our paper explores the automatic annotation of time-series charts with additional contextual information, sourced from an external news headlines corpus. This work builds on prior work that explores annotation techniques for charts and corpus-driven approaches for enriching data narratives.

%Our paper builds on prior work on chart annotation. %annotating charts as well as corpus-driven methods for enriching data narratives.
% \maneesh{My sense is that we should significantly shorten the related work -- keep all the citations but use more general sentences that can have many citations at the ends of them. We should also incorporate any papers the reviewers mentioned or anything new we know about now. }\vidya{We need to prune citations as we are only allowed one page of them}
%\subsection{Annotation Techniques for Charts}
%Authors add annotations to charts to guide a reader's attention to visual features in the chart, explaining what the underlying data means and providing additional context~\cite{segel:2010,hullman:2011}. %
Prior research has explored how authors add annotations to charts guiding a reader's attention to visual features in the chart and explaining what the underlying data means with additional context~\cite{segel:2010,hullman:2011}. 
Kong and Agrawala developed techniques for analyzing charts to recover visually salient features of the data-encoding marks (e.g., min, max, mean values). Users could then interactively add graphical and text annotations or overlays to facilitate chart reading~\cite{kong2012graphical,kong09}.

Kandogan~\cite{kandogan:2012} introduced just-in-time descriptive analytics by employing statistics to generate annotations for clusters and outliers. Autotator is a semi-automatic chart annotation system that provides suggestions for three annotation tasks, i.e., (1) labeling a chart type, (2) annotating bounding boxes, and (3) associating a quantity~\cite{Autotator:2019}. Our work focuses on providing external context rather than just annotating a chart with low-level statistical information. Several projects have also aimed to automatically generate visualizations for a text article~\cite{Gao2014} and  personalize articles with text and visualizations based on information about the specific reader~\cite{Adar2017,Kim:2016,Alhalaseh:2018,2020-comm-interactive-articles}.%, for example, automatically generates thematic maps to accompany an input news article, using NLP and entity extraction techniques to first identify quantities and geo-locating entities mentioned in the article and then select from a corpus of external data sets to serve as the underlying data for the map. Similarly, researchers have developed techniques to personalize new articles with text notes and visualizations based on information about the specific reader~\cite{Adar2017,Kim:2016,Alhalaseh:2018,2020-comm-interactive-articles}. %\maneesh{Are there other personalization papers we need to reference?}

% \maneesh{This paragraph probably needs to stay longer and also potentially updated to highlight our new contributions beyond their work. Update to cover beyond NYT in our work.}
Closest to our work is Contextifier~\cite{Hullman2013}, which uses news headlines to provide external contextual annotations for line charts. Their implementation takes as input, a financial news article and generates an annotated price chart for the most frequently mentioned stock in the article. They consider linguistic relevance, the number of article views, and the visual saliency of chart peaks to identify the headlines and chart features to annotate. But their implementation is limited to a relatively small corpus of stocks and financial articles covering 2010-2012. In contrast, by working with a large corpus of 2,425,425 NYT headlines from 1980 to 2022, our API can handle more general data sets and covers a longer period of time. Finally, unlike our work, Contextifier does not provide a JavaScript API that can be incorporated into web-based visualizations.

.

\section{\toolname{} JavaScript API}
The \toolname{} API comprises two main components: a prominence feature detector (Section~\ref{sec:featuredetector}) and a contextual annotation recommender (Section~\ref{sec:annotationrecommender}). While both components are exposed to chart authors as JavaScript function calls {\ttfamily getChartFeatures} and {\ttfamily getAnnotations} respectively, the prominence feature detector is implemented in JavaScript and the annotation recommender is implemented as a Python REST API. We have architected the two components to be completely independent of one another to facilitate reuse in contexts beyond chart annotation. The primary input to both components is a time-series data set $T$, an associated granularity $T_G$ (e.g., year, month, day) that specifies the sample rate of the time series, and a set of keywords, $T_W$ that describe the overall data domain of the chart (e.g., ``california wildfires'', ``covid cases'').

\toolname's recommender component includes a built-in annotation database consisting of NYT headlines that we curated using the NYT Archive API~\cite{NYTAPI}. We considered this as our built-in database because the NYT is a general and broad news source covering a wide variety of events and topics and hence would be more likely to provide relevant annotation for a variety of time-series data sets. We used the NYT API to obtain all of the metadata, including the headline, publish date, article type (e.g., news, opinion), and lede paragraph for NYT articles published between January 1, 1980 and December 31, 2022. We removed duplicates and filtered the database to include only articles that cover news events. The resulting database contains metadata for 2,425,425 articles.

\subsection{Prominence Feature Detector}
\label{sec:featuredetector}
\toolname's prominence feature detector takes a time-series data set as input and outputs a ranked ordering of the peaks (or valleys) in the data based on their prominence.
%---i.e. the most prominent peak (or valley) is ranked first. 
In JavaScript, a chart author invokes the \toolname{} feature detector as, 
\begin{equation*}
    P = \texttt {getChartFeatures}(T)
\end{equation*}
where $T$ is a time-series data set, $P$ is a 
set of peak (or valley features $p_i$ and $i$ is the prominence rank of feature $p_i$. The input time series $T$ is associated with a granularity $T_G$ (e.g. year, month, day) that specifies the sample rate of the time series. 
Each output feature $p_i$ can be a single point in time (corresponding to a peak or valley) or interval in time (corresponding to a trend or slope).

%\input{figures/persistence.tex}

% \begin{figure*}[ht]
% %\vspace{-1em}
%    % \includegraphics[width=\textwidth]{figures/resultsPt1.pdf}
%       \includegraphics[width=\textwidth]{figures/resultsV2/png/presidentialApprovalV4.png}
%     \vspace{-2em}
%     \caption{Chart depicting times-series data of Presidential Approval Ratings. We used the \toolname{} API to identify prominence features (i.e., peaks and valleys) and to recommend NYT headlines for them. All annotations shown are the top-ranked.}% Note that some of the peaks and valleys \toolname{} deemed prominent, are unlabeled because the annotation recommender did not find a suitable headline for them.}
%     \label{fig:resultsPt1}
% \end{figure*}

\begin{figure*}[ht]
    %\vspace{-1em}
    \includegraphics[width=\textwidth]{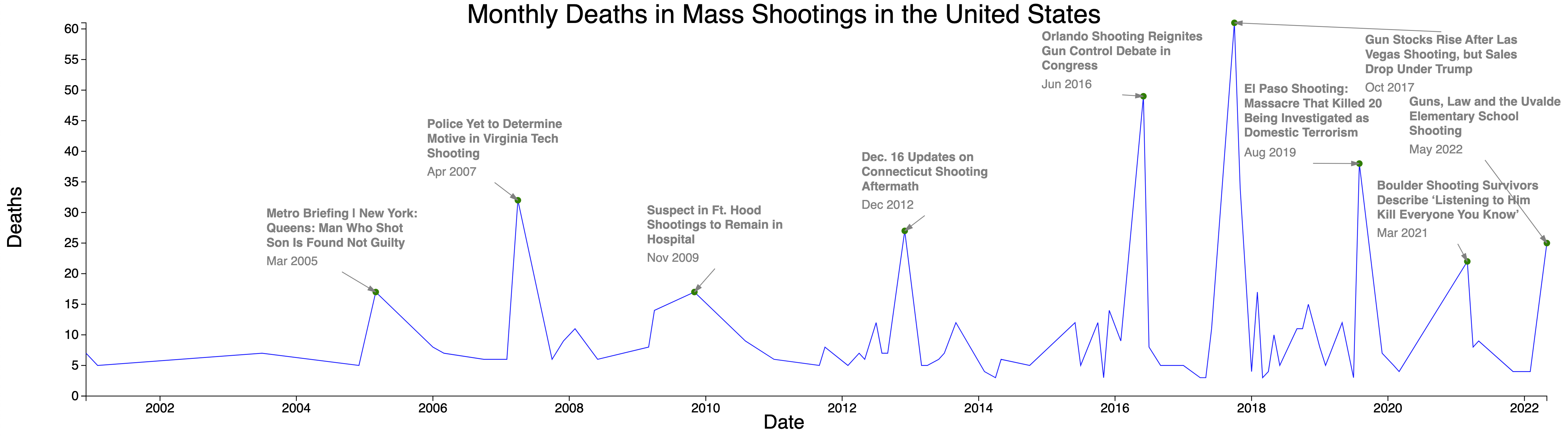}
    \includegraphics[width=\textwidth]{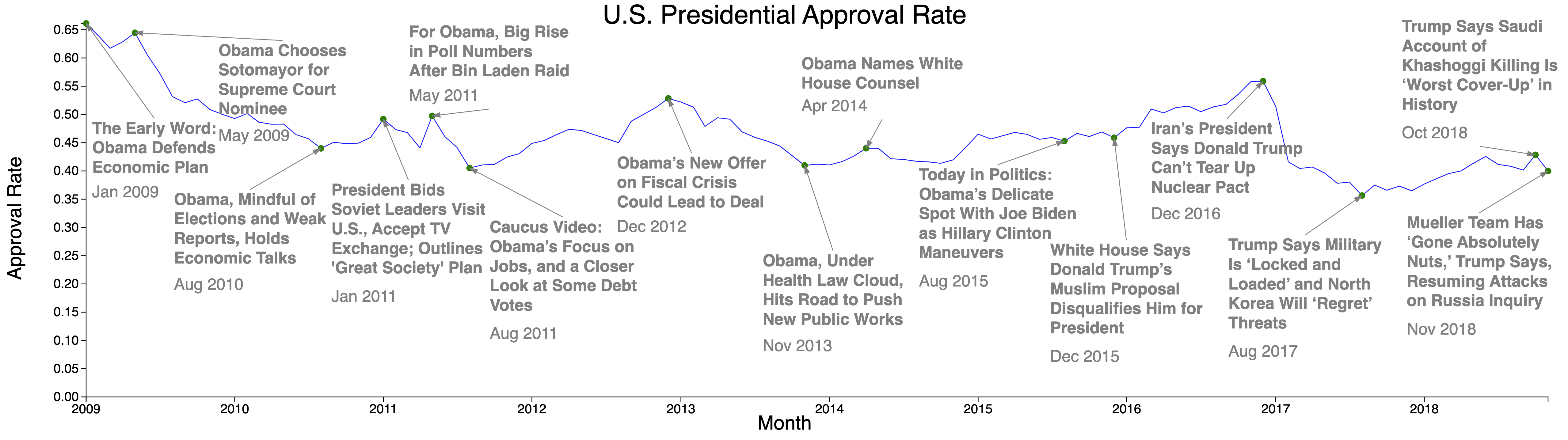}
    \vspace{-2em}
    % \caption{Charts depicting times-series data of Mass Shooting Deaths (a), Unemployment Rate (b), and Presidential Approval Ratings (c). We used the \toolname{} API to identify prominence features (peaks for a, peaks and valleys for b and c) and to recommend NYT headlines for them. All annotations shown are the top-ranked. Note that some of the peaks and valleys \toolname{} deemed prominent, are unlabeled because the annotation recommender did not find a suitable headline for them.}
        \caption{Charts depicting times-series data of Mass Shooting Deaths (a) and Presidential Approval Ratings (b). \toolname{} API identifies prominence features (peaks for a, peaks and valleys for b) and recommends NYT headlines. All annotations shown are the top-ranked. Note that some of the peaks and valleys \toolname{} deemed prominent, are unlabeled because the annotation recommender did not find a suitable headline.}
    \label{fig:resultsPt1}
\end{figure*}

% \begin{figure*}
% \vspace{-1em}
%   %  \includegraphics[width=\textwidth]{figures/resultsPt2.pdf}
%    \includegraphics[width=\textwidth]{figures/resultsV2/png/simpsons.png}
%     \vspace{-2em}
%     \caption{Chart depicting times-series data of Simpsons Viewership. We used the \toolname{} API to identify prominence features and to recommend NYT headlines for them. All annotations shown are the top-ranked. Note that some of the peaks \toolname{} deemed prominent, are unlabeled because the annotation recommender did not find a suitable headline for them.}
%     \label{fig:resultsPt2}
% \end{figure*}

% \begin{figure}
% \vspace{-1em}
%     \includegraphics[width=\columnwidth]{figures/resultsV2/png/tvnews_afghanistan.png}
%     \includegraphics[width=\columnwidth]{figures/resultsV3/png/covid.png}
%     \includegraphics[width=\columnwidth]{figures/resultsV3/png/simpsons.png}
%     \vspace{-2em}
%     \caption{Charts depicting times-series data of Cable TV News coverage of Afghanistan (a), COVID-19 Cases (b), and Simpsons Viewership (c). We used the \toolname{} API to identify prominence features (peaks a, b, and c) and to recommend NYT headlines for them. All annotations shown are the top-ranked. Note that some of the peaks \toolname{} deemed prominent, are unlabeled because the annotation recommender did not find a suitable headline for them.}
%     \label{fig:resultsPt2}
% \end{figure}

\begin{figure*}[ht]
\vspace{-1em}
    \includegraphics[width=\textwidth]{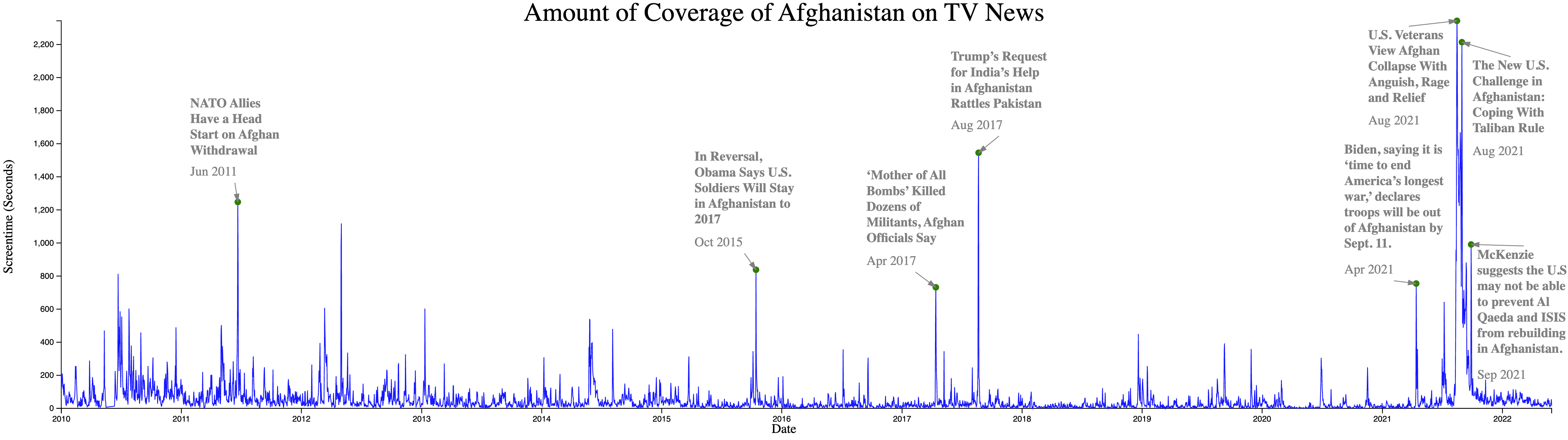}
    \includegraphics[width=\textwidth]{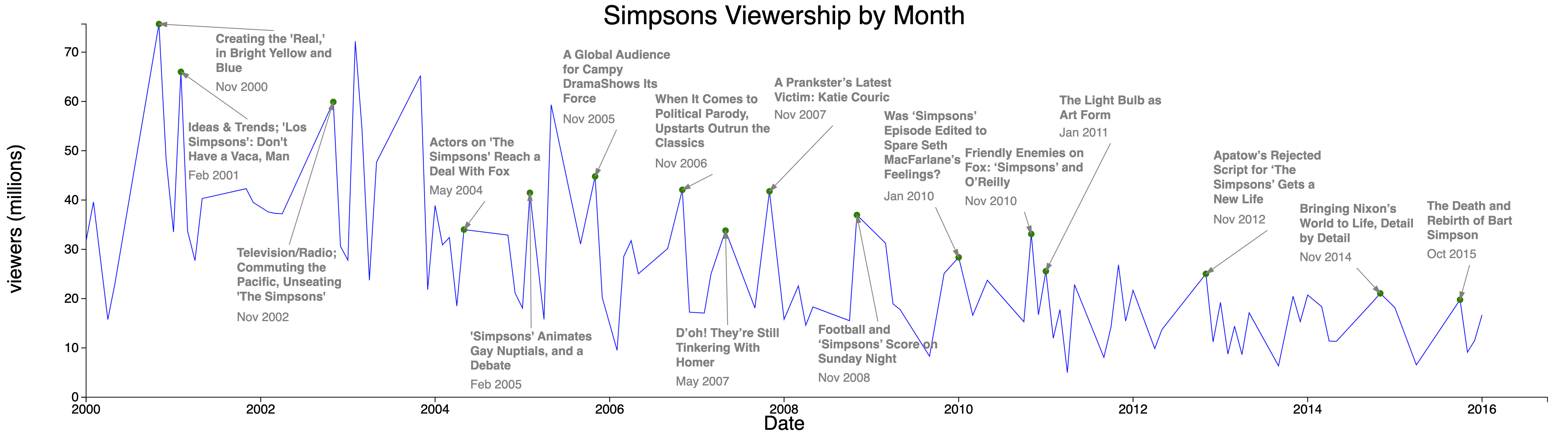}
    \vspace{-2em}
    \caption{Charts depicting times-series data of Cable TV News coverage of Afghanistan (a) and Simpsons Viewership (b). We used the \toolname{} API to identify prominence features (peaks) and to recommend NYT headlines for them. All annotations shown are the top-ranked. Note that some of the peaks \toolname{} deemed prominent are unlabeled because the annotation recommender did not find a suitable headline for them.}
    \label{fig:resultsPt2}
\end{figure*}

\toolname{} provides a feature detector that measures the prominence of peaks (or valleys) in the data as a proxy for their visual salience. The detector computes the prominence of peaks using a topological data analysis technique~\cite{Huber2020,Kirmse2017}, which uses a watershed model to determine how long a peak remains independent when the surrounding relief is metaphorically drained of water. % (Figure~\ref{fig:persistence}).

\subsection{Contextual Annotation Recommender}
\label{sec:annotationrecommender}

\toolname's contextual annotation recommender uses keyword anomalies to surface relevant headlines. The keywords appear within a time range associated with a visual feature but do not appear in ranges associated with other visual features, allowing us to identify keywords associated with events, often relating to cause or effect. While there are multiple methods of determining relevancy, \toolname{} implements relevance as a time-based variant of Term Frequency - Inverse Document Frequency (TF-IDF)~\cite{Manning2009}. The API finds annotations that are most uniquely relevant to a visually prominent chart feature $p$ compared to a set of other chart features $Q$.

The input parameters to the contextual annotation recommender are (1) 
a visually prominent feature $p$, (2) a set of other features $Q$, and 
(3) the time-series data $T$. The annotation recommender outputs a ranked set of contextual headline annotations $H$ for feature $p$. %In \toolname{} syntax,
\begin{equation*}
    H = \texttt {getAnnotations}(p, Q, T)
\end{equation*}
to obtain a ranked set of headlines $H$ for feature $p$, where for each headline $h_i \in H$, $i$ is its relevance rank. Note that $p$ and $Q$ can consist of features identified by \toolname's visual feature detector (Section~\ref{sec:featuredetector}) or may be specified manually by the chart author.

\toolname's annotation recommender is designed to find 
NYT headlines that are most uniquely relevant to feature $p$ compared to the other features in $Q$. The finest granularity of publication date for our NYT headlines is the day of publication. So, the annotation recommender starts by constructing a time range $r$ in days for each feature in $\{p\} \cup Q$, using the sample frequency granularity $T_G$ associated with the input time-series data $T$. The set of such time ranges is $R$, and $r_p \in R$ is the range associated with feature $p$.

Next, for each feature time range $r \in R$, the recommender filters the annotation database to the NYT articles published within the range and includes headlines or lede paragraphs containing descriptive keywords $T_W$ associated with the time-series data $T$. For each feature time range $r$, we obtain a set of headlines $H_r$ that are generally relevant to the time-series data. The recommender then applies TF-IDF \cite{Manning2009} to score each headline $h \in H_{r_p}$ corresponding to feature $p$ by summing the TF-IDF score for each word.

To apply TF-IDF in this context, the recommender treats the set of headlines $H_r$ associated with each feature time range $r$ as a {\em document}, while the {\em corpus} consists of the set of documents  $\mathcal{H}_R = \bigcup_{r \in R} H_r$ corresponding to all of the feature time ranges. 
Specifically, 
\toolname{} scores each headline $h \in H_{r_p}$ corresponding to $p$ by summing the TF-IDF score for each of its words; i.e., for each word (or term) $t$ in such a headline, the term frequency $\tf$ is computed as:
\begin{equation}
 \tf{(t,H_{r_p})} = \frac{f_{t,H_{r_p}}}{\sum_{t' \in H_{r_p}}f_{t',H_{r_p}}}.
 \label{eq:tf}
\end{equation}
This measures the relative frequency $t$ that appears in the headlines corresponding to $p$. The denominator is the number of terms that appear in these headlines. The inverse document frequency, $\idf$ for term $t$ is computed as:
\begin{equation}
\label{eq:idf}
\idf{(t,\mathcal{H}_R)} = \log \left( \frac{|\mathcal{H}_R|}{|\{H_r \in \mathcal{H}_R : t \in H_r\}|} \right)
\end{equation}
where the numerator is the total number of documents. Since there is one document (or set of headlines) for each feature in $\{p\} \cup Q$, the numerator is equivalent to the number of features in $\{p\} \cup Q$. The denominator is the number of documents that contain the term $t$. The final TF-IDF score for $t$ is then: 
\begin{equation}
\label{eq:tfidf}
\tfidf{(t,H_{r_p},\mathcal{H}_R)} = \tf{(t,H_{r_p})} \cdot \idf{(t,\mathcal{H}_R)}.
\end{equation}
\toolname{} treats the TF-IDF score for each headline $h \in H_{r_p}$ as its rank, with the highest scoring headline ranked first, and returns the ranked set of headlines as the recommended annotations $H$ for $p$.

\subsection{Using the \toolname{} API}
\label{sec:usingAPI}
\toolname{} can be applied to any chart that is developed using web-based tools, including D3.js~\cite{Bostock2011} and Observable notebooks (\url{https://observablehq.com/@ibnzterrell/almanac-observable-demo}). Figure~\ref{fig:usingAPI} illustrates how a chart author might apply the \toolname{} API to construct an annotated chart.  We have also experimented with incorporating \toolname{} as a dashboard extension to annotate Tableau charts~\cite{tableauextension}. In the context of Tableau, users can either manually click on prominent features in the chart or rely on \toolname{} to automatically identify features prior to displaying the recommended annotations. \toolname{} is publicly available on Github (\url{https://github.com/ibnzterrell/Almanac}) and a video showing these various applications using the API can be found in the supplemental material.

\section{Results and Future Work}
We demonstrate the utility of the \toolname~API with various D3.js charts displaying the top-ranked annotations recommended by \toolname, as shown in Figures~\ref{fig:teaser},~\ref{fig:resultsPt1}, and~\ref{fig:resultsPt2}. In all of these cases, the API was used to identify peaks (and in some cases, also identify valleys) and then generate annotation recommendations for those features. In all cases, we show the default, top-ranked annotation headline returned by the annotation recommender. However, if the recommender cannot find a suitable headline, the feature initially goes unlabeled. Unlabeled features can then be manually labeled. These examples show how the headlines annotations can provide useful context for prominent features in charts depicting various types of data.  Additional annotated line chart examples and datasets can be found in the supplementary material. 

Figure~\ref{fig:teaser} shows how the top-ranked annotations name California wildfires~\cite{Ares2020Wildfires}. Prominent features in the chart are annotated with text containing relevant names and locations of these wildfires, providing useful context in understanding the effects of the fires. For example, the highest peak on July 2018 displays the annotation, \textit{``Carr Fire in California Claims Another Victim, Bringing Death Toll to 6.''} Other recommended annotations for the various prominent features in the chart include, \textit{``Wildfire Near Yosemite Destroys Berkeley's Family Campground''} in July 2013 and \textit{``California ``Wildfires Updates: 48 Dead in Camp Fire, Toll Expected to Rise,''} referring to the deadly  fire that occurred in August 2018.

Similarly, \toolname~provides headlines context related to the location and issues surrounding the mass shootings in Figure~\ref{fig:resultsPt1}a. The annotation, \textit{``Police Yet to Determine Motive in Virginia Tech Shooting'' } for example, indicates that a mass shooting took place at Virginia Tech in April 2007. The peak corresponding to the Orlando shooting in 2016 displays the annotation, \textit{``Orlando Shooting Reignites Gun Control Debate in Congress,''} naming the location and some of the national concerns surrounding it.

Figure~\ref{fig:resultsPt1}b shows a chart example of the U.S. presidential approval rating between the years 2009 and 2019~\cite{Gallup2018PresidentialApproval}. \toolname{} provides causal explanations for the reader to better understand the context of the fluctuations in presidential approval ratings. For example, a peak in President Obama's approval rating on May 26, 2008, shows an annotation, \textit{``Obama Chooses Sotomayor for Supreme Court Nominee,''} indicating that perhaps the public viewed the nomination of the first Hispanic woman to the Supreme Court positively. Another peak annotated with, \textit{``For Obama, Big Rise in Poll Numbers After Bin Laden Raid''} suggests that the event was viewed favorably by the public, and Obama's approval ratings went up.

%%%%%%%%%%%%%%%%%%%%%%%%%%%%%%%%%%%%%%%%%%%%%%%%%%%

Figure~\ref{fig:resultsPt2}a shows the amount of coverage of Afghanistan on cable TV news channels between 2010 and 2022~\cite{Hong2021}. The annotation for the peak in Apr 2021 \textit{``Biden saying it is `time to end America's longest war,' declares troops will be out of Afghanistan by Sept. 11''} explains that the U.S. is planning to leave the country, which gives context for the peak in coverage of the country. %The next labeled peak is the global peak in coverage of Afghanistan and the annotation \textit{``U.S. Veterans View Afghan Collapse With Anguish, Rage and Relief''} explains that the Afghan government collapsed -- presumably after the U.S. withdrawal suggested by the previous annotation.
There are cases, however, where the API generates annotations that may not be very relevant to the events occurring at these prominent features. For example, the recommended annotation for Aug 22, 2017, \textit{``Trump’s Request for India’s Help in Afghanistan Rattles Pakistan''} is only tangentially relevant to the events occurring in that country.

% Figure~\ref{fig:resultsPt2}b shows daily COVID-19 cases in the United States from March 2020 to April 2022~\cite{NYT2022Covid}. \toolname{} recommends annotations that describe reactions and responses to temporal events. For example, the highest peak in COVID-19 cases on January 12, 2022, says that `\textit{`More countries shift to self-administered antigen tests in a race to stop Omicron''}. %The annotation indicates that a sharp rise in cases perhaps resulted in additional measures that countries adopted to cope with the pandemic. The annotation for the spike on January 2021, \textit{``The U.S. vaccine rollout needs to ‘hit the reset,’ an ex-F.D.A. chief says''} is a relevant annotation to describe the new strategy the government had to adopt to get vaccines out to as many people as possible. However, because \toolname{} uses headlines from the NYT, they may be skewed towards events in New York.
%the text can sometimes appear in a news- or journalist-style of writing. Another limitation of the API is that some of the headlines in the NYT database are skewed towards news events in New York. 
%For example, a spike in the chart for January 26, 2022, is annotated with, \textit{``New York City schools will shorten isolation to 5 days for most students who test positive.''} While relevant, it is not a general description of the spike concerning case numbers across the United States.

Automatic annotations can also be useful for understanding information pertaining to an attribute of interest. Figure~\ref{fig:resultsPt2}b shows data on viewership of {\em The Simpsons} TV show~\cite{Brown2022Simpsons}. The 
annotations provide information related to the show. The annotation, \textit{``Actors on `The Simpsons' Reach a Deal With Fox''} explains that in May 2004, efforts by actors on the show to win new contracts finally came through just as the viewership was steadily declining. An annotation for a spike in viewership for November 2007, \textit{``Ratings: Sunday Means Cartoons (and `60 Minutes')''} indicates that the show was a popular animated comedy line-up during the Sunday TV program.

While \toolname{} provides an API for annotating time-series charts with NYT headlines for a variety of examples, future work should consider incorporating other annotation sources such as Google search results, Wikipedia Events~\cite{Wikipedia2022Events}, or domain-specific corpora. Recent work has explored the problem of rendering charts at different sizes while preserving their prominent features~\cite{semanticresizing:2021}. \toolname's prominence feature detector command could be used in automating responsive chart design and for different target displays. \toolname's annotation recommender not only identifies relevant headlines but also scores those headlines and the terms using our time-based variant of TF-IDF. This metadata could be useful to find additional articles related to the event topic and thereby support the creation of data narratives beyond annotation recommendations. 

%% Vidya: Commented out the evaluation as I'm not convinced it's critical for a four-pager. Maybe we have a modified version.
% \section{Evaluation}
% \input{sections/05_evaluation}

% \section{Limitations and Future Work}
% \label{sec:discussion}
% \input{sections/06_futurework}

\section{Conclusion}
To summarize, we introduce \toolname, a JavaScript API that identifies visually prominent features in a time-series chart and recommends annotations sourced from the NYT corpus of news headlines to provide additional contextual information to the prominent features. 
%The tool consists of a prominence feature detector and a contextual headline finder, allowing for flexibility in how the prominent features are chosen and annotated in the chart. 
We demonstrate the utility of \toolname{} by annotating various D3.js and Vega-Lite time-series charts from a variety of data sources. 
%The annotations provide additional information that describes various prominent events occurring in the chart. 
% A web survey designed to assess how readers perceive the relevance of the annotations recommended by \toolname, indicates that the top-ranked annotation provides relevant information describing a given prominent feature in the chart. 
By serving as an API-level tool that can be integrated with other JavaScript libraries, \toolname{} can facilitate the creation of well-annotated charts across the Web.

% Force bib to end
%\clearpage

%\bibliographystyle{abbrv}
\bibliographystyle{abbrv-doi}

\bibliography{references}
\end{document}